\documentclass[]{spie}  

\usepackage{amsmath,amsfonts,amssymb}
\usepackage{graphicx}
\usepackage[colorlinks=true, allcolors=blue]{hyperref}
\usepackage{pifont}
\usepackage{multirow}
\usepackage{array}
\usepackage{subcaption}
\newcolumntype{L}[1]{>{\raggedright\let\newline\\\arraybackslash\hspace{0pt}}m{#1}}
\newcolumntype{C}[1]{>{\centering\let\newline\\\arraybackslash\hspace{0pt}}m{#1}}
\newcolumntype{R}[1]{>{\raggedleft\let\newline\\\arraybackslash\hspace{0pt}}m{#1}}

\newcommand{\cmark}{\ding{51}}%
\newcommand{\xmark}{\ding{55}}%

\makeatletter
\def\thickhline{%
  \noalign{\ifnum0=`}\fi\hrule \@height \thickarrayrulewidth \futurelet
   \reserved@a\@xthickhline}
\def\@xthickhline{\ifx\reserved@a\thickhline
               \vskip\doublerulesep
               \vskip-\thickarrayrulewidth
             \fi
      \ifnum0=`{\fi}}
\makeatother

\newlength{\thickarrayrulewidth}
\title{A ResNet is All You Need? Modeling A Strong Baseline for Detecting~Referable~Diabetic~Retinopathy in Fundus Images}

\author[a]{Tom\'as~Castilla}
\author[b]{Marcela~S.~Mart\'inez}
\author[c]{Mercedes~Legu\'ia}
\author[a,d]{Ignacio~Larrabide}
\author[a,d]{Jos\'e~Ignacio~Orlando}

\affil[a]{Yatiris Group, PLADEMA Institute, UNICEN, Campus Universitario, Tandil, Argentina}
\affil[b]{Centro de Oftalmolog\'ia Mart\'inez, Hip\'olito Yrigoyen 380, Pehuaj\'o, Argentina}
\affil[c]{Servicio de Oftalmolog\'ia, Hospital de Alta Complejidad En Red "El Cruce" Dr. N\'estor Carlos Kirchner, Av. Calchaqu\'i 5401, Florencio Varela, Argentina}
\affil[d]{Consejo Nacional de Investigaciones Cient\'ificas y T\'ecnicas, CONICET, Tandil, Argentina}

\authorinfo{Further author information: (Send correspondence to J.I.O.)\\J.I.O.: E-mail: jiorlando@pladema.exa.unicen.edu.ar, Telephone: +54 249 438 5690}

\begin{document} 
\maketitle

\begin{abstract}
Deep learning is currently the state-of-the-art for automated detection of referable diabetic retinopathy (DR) from color fundus photographs (CFP).
While the general interest is put on improving results through methodological innovations, it is not clear how good these approaches perform compared to standard deep classification models trained with the appropriate settings.
In this paper we propose to model a strong baseline for this task based on a simple and standard ResNet-18 architecture.
To this end, we built on top of prior art by training the model with a standard preprocessing strategy but using images from several public sources and an empirically calibrated data augmentation setting. To evaluate its performance, we covered multiple clinically relevant perspectives, including image and patient level DR screening, discriminating responses by input quality and DR grade, assessing model uncertainties and analyzing its results in a qualitative manner.
With no other methodological innovation than a carefully designed training, our ResNet model achieved an AUC = 0.955 (0.953 - 0.956) on a combined test set of 61007 test images from different public datasets, which is in line or even better than what other more complex deep learning models reported in the literature.
Similar AUC values were obtained in 480 images from two separate in-house databases specially prepared for this study, which emphasize its generalization ability.
This confirms that standard networks can still be strong baselines for this task if properly trained.
\end{abstract}


\section{INTRODUCTION}
\label{sec:intro}  

Automated detection of referable diabetic retinopathy (DR) in color fundus photographs (CFP) has been extensively explored in the literature~\cite{grzybowski2020artificial}, as it aids in reducing the cost of large scale screening campaigns and favors the access to early treatment~\cite{scotland2010costs,gulshan2016development,grzybowski2020artificial}. Deep Convolutional Neural Networks (CNNs) have became the de facto standard for this task~\cite{pires2015beyond,van2016fast,gargeya2017automated,quellec2017deep,li2019diagnostic,grzybowski2020artificial,yip2020technical,zago2020diabetic,huang2021identifying,hervella2022multimodal}, improving results up to a point in which they are either comparable or superior to those obtained by trained ophthalmologists.~\cite{gulshan2016development,gulshan2019performance, grzybowski2020artificial}
Authors generally attribute the accuracy of their results to their proposed methodological innovations, which can vary from novel architectural modules~\cite{liu2019referable} to new pre-training strategies~\cite{quellec2017deep,pires2019data,hervella2022multimodal} or multitask  designs~\cite{zhou2018fundus}, among others~\cite{grzybowski2020artificial,islam2020deep}.
Surprisingly, we observed that little to no attention is payed to simple yet influential factors such as integrating multiple data sources for training, applying good preprocessing strategies on the input images or calibrating data augmentation techniques. \textit{Our hypothesis is that this leads to incomplete or unfair comparisons with suboptimally trained baselines}, being unknown if training standard classification networks in a proper manner is more beneficial than using complex architectures. 

Recent studies have analyze the influence of these factors on DR classification or grading algorithms~\cite{yip2020technical,huang2021identifying}. In particular, Huang \textit{et al.}~\cite{huang2021identifying} demonstrated that a systematic investigation of training losses, input image resolutions, data augmentation strategies and learning rate scheduling techniques allows a standard ResNet-50 model to achieve state-of-the-art results for DR grading. Building on top of these results, in this paper we present a strong baseline for detecting referable DR from CFPs based solely on a properly trained standard ResNet architecture. This binary classification task differs in that, instead of classifying the DR grade, it detect cases that need to be referred to an ophthalmologist (namely moderate to severe non-proliferative DR and proliferative DR) from those that are not at risk (with no DR or mild DR signs). Our model is based on a simple ResNet-18 architecture, pre-trained on ImageNet and fine-tuned using images preprocessed with standard practices. To increase its ability to generalize to new databases, we crafted a multiethnic training set based on several public sources, while incorporating a series of data augmentation strategies empirically calibrated using a held-out validation set. The proposed model was evaluated using nine public test sets with more than 61.000 images, and two additional in-house data sets specially prepared for this study, namely Martínez and HEC. To this end, we applied an extensive protocol accounting for multiple clinically relevant observations, including image level and patient level evaluations, discriminating responses according to the quality and DR grade of the input, assessing model uncertainties and qualitatively analyzing model's attention through class activation maps. Notoriously, our simple ResNet-18 baseline performed in line or even better than other recently published, more complex deep learning architectures, even though no problem-specific innovations were implemented in ours. These results empirically confirm our hypothesis that properly trained standard classification networks can be strong enough to compete with other state-of-the-art approaches, and that existing comparisons in the literature are sometimes unfair and suboptimal. To avoid this issue and favor reproducibility, we publicly release our training, validation and test data partitions and our test set predictions. Finally, we identify a series of weak points in this baseline and recommend alternative lines of research to address them.



\section{MATERIALS AND METHODS}
\label{sec:methods}

\subsection{Materials}
\label{subsec:materials}

We used public databases that provided either referable/non-referable DR labels or DR grades or manual segmentations of DR lesions (Table~\ref{tab:datasets}). 
Binary labels were generated from grades by mapping no DR or mild non-proliferative DR (NPDR) to the non-referable DR class, and moderate and severe NPDR or proliferative DR (PDR) to the referable DR class. 
When only lesion segmentations were available, we assigned binary labels using Decenciere's \textit{et al.} criterion\cite{decenciere2014feedback}. 
We also created two in-house private test databases by retrospectively collecting anonymized images from the Ophthalmological Center Martínez (Pehuajó, Argentina)--Martínez set--and Hospital “El Cruce” (Florencio Varela, Argentina)--HEC set--. 
The protocol to access the images was approved by the hospital's Ethics Committee, in accordance to the tenets of the Declaration of Helsinki. 
Martínez comprises 484 images acquired using a non-mydriatic Cristal Vue NFC-700 fundus camera with a field-of-view (FOV) of $45^\circ$. 
HEC, on the other hand, corresponds to an atlas of 35 images acquired using a Topcon TRC-NW8 non-mydriatic camera, similar to the one used in the DR2 set. 
Both sets were manually labeled by two independent ophthalmologists (MM and ML, respectively), indicating which images correspond to referable DR cases.

\subsection{Image preprocessing}
\label{subsec:methods-preprocessing}

Referable DR classification approaches use different image preprocessing strategies, with two being the most commonly seen, namely cropping out the empty areas outside the CFP field-of-view (FOV),\cite{sahlsten2019deep,liu2019referable,li2022deep} and enhancing the contrast of DR lesions and vascular structures by subtracting an estimated background.~\cite{graham2015kaggle,van2016fast,huang2021identifying} 

\begin{figure}
    \centering
    
    \begin{subfigure}[]{0.245\textwidth}
      \includegraphics[width=\textwidth]{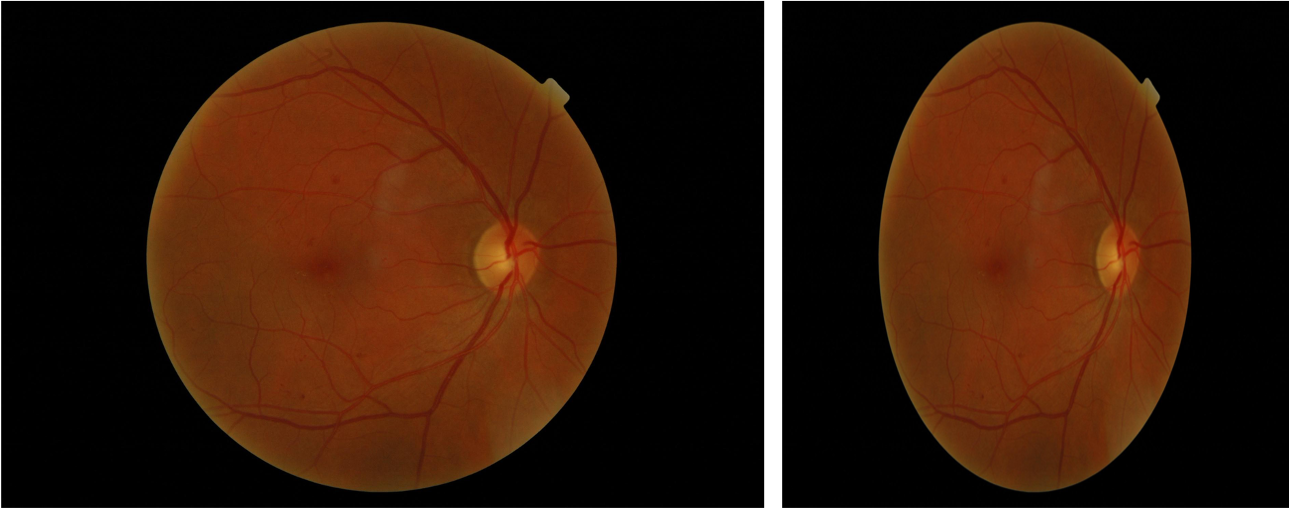}
      \caption{\label{fig:prepro-bad} No preprocessing}
    \end{subfigure}
    \begin{subfigure}[]{0.245\textwidth}
      \includegraphics[width=\textwidth]{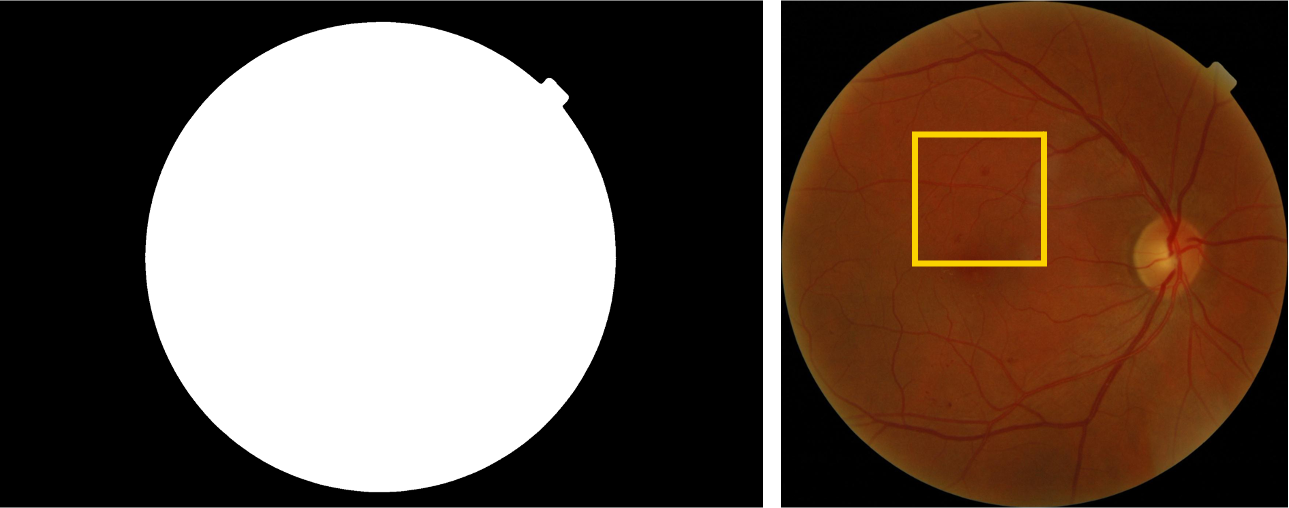}
      \caption{\label{fig:prepro-first} FOV crop}
    \end{subfigure}
    \begin{subfigure}[]{0.29\textwidth}
      \includegraphics[width=\textwidth]{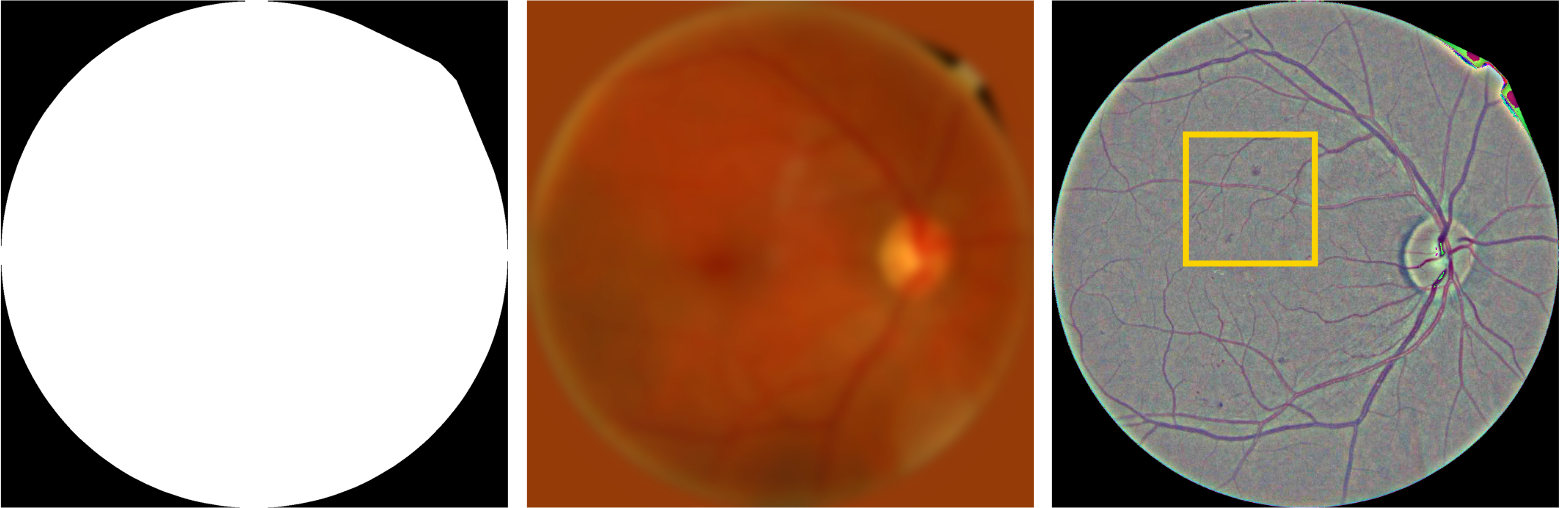}
      \caption{\label{fig:prepro-second} Contrast enhancement}
    \end{subfigure}
    \begin{subfigure}[]{0.19\textwidth}
      \includegraphics[width=\textwidth]{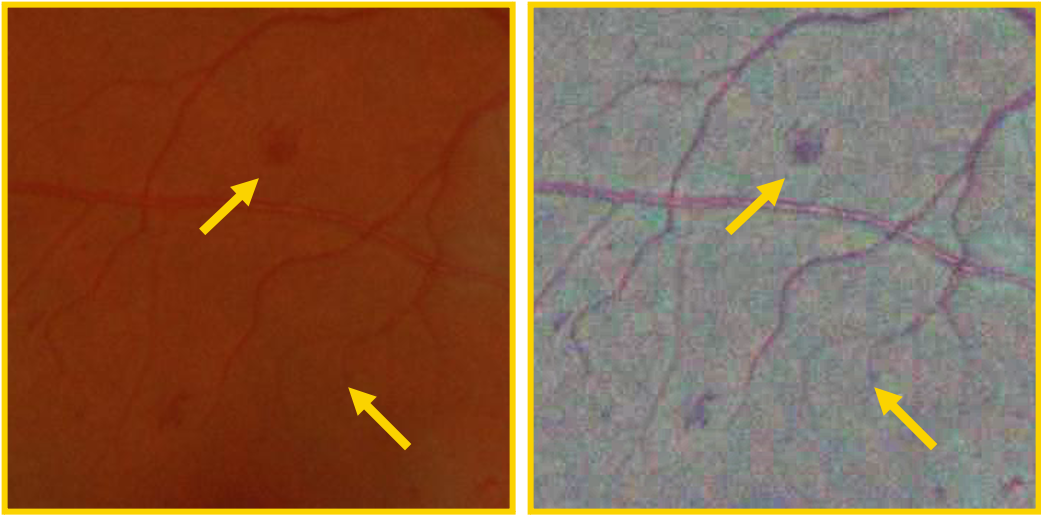}
      \caption{\label{fig:prepro-detail} Detail}
    \end{subfigure}
    \vspace{0.3cm}
    \caption{\label{fig:preprocessing} Preprocessing operations. (a) If no preprocessing is used, image resizing affects its aspect ratio. (b) FOV cropping better preserves the aspect ratio. Left: Estimated FOV. Right: Image cropped around the FOV and resized to $512 \times 512$ pixels. (c) Contrast enhancement~\cite{graham2015kaggle}. Left: convex hull of the estimated FOV. Center: Estimated background. Right: contrast enhanced image $I_\text{ce}$. (d) Details of input image $I$ (left) and $I_\text{ce}$ (right).}
    
\end{figure}

FOV cropping allows to standardize the resolution of images that were collected from multiple sources or acquired with different devices. Furthermore, it preserves their overall aspect ratio when resized for training, as seen in Figures~\ref{fig:prepro-bad} and \ref{fig:prepro-first}. This operation requires to approximate the border of the FOV, which we did by summing the red, green and blue intensities of the input image in a pixel-wise manner, then thresholding the resulting matrix using Otsu's method, and finally applying a filling holes operation. Figure~\ref{fig:prepro-first} (left) depicts the resulting binary mask. The smallest possible bounding box was then predicted from it, and its coordinates are used to crop the area of interest before resizing. The final image is observed in Figure~\ref{fig:prepro-first} (right). 

\begin{table}[]
    \centering
    \caption{Training, validation and test partitions created from public databases and our two in-house sets.}
    \vspace{0.2cm}
    \resizebox{0.9\textwidth}{!}{
    \begin{tabular}{ C{3.5cm} | C{3.4cm} | C{3.4cm} | C{1.5cm} | C{1.8cm} | C{1.8cm} | C{1.8cm} }
        \thickhline
        \multirow{2}{3.5cm}{\centering \textbf{Dataset}} & \multicolumn{3}{c|}{\textbf{Num. samples}} & \multirow{2}{1.8cm}{\centering \textbf{Training}} & \multirow{2}{1.8cm}{\centering \textbf{Validation}} & \multirow{2}{1.8cm}{\centering \textbf{Test}} \\
        \cline{2-4}
                & \textbf{Non-referable DR} & \textbf{Referable DR} & \textbf{Total} & & & \\
        \thickhline
        APTOS2019~\footnote[1] & 2175 & 1487 & 3662 & 3662 & 0 & 0 \\
        \hline
        DeepDRID \cite{liu2022deepdrid} & 900 & 700 & 1600 & 1200 & 0 & 400 \\
        \hline
        DDR \cite{li2019diagnostic} & 6896 & 5626 & 12522 & 6260 & 2503 & 3759 \\
        \hline
        EyePACS~\footnote[2] & 71548 & 17154 & 88702 & 28098 & 7026 & 53576 \\
        \hline
        IDRiD~\cite{porwal2018indian} & 193 & 323 & 516 & 372 & 40 & 103 \\ 
        \hline
        FCM-UNA~\cite{benitez2021dataset} & 191 & 566 & 757 & 0 & 0 & 757 \\
        \hline
        1000Fundus~\cite{cen2021automatic} & 56 & 88 & 144 & 0 & 0 & 144 \\
        \hline
        DR2~\cite{pires2015beyond} & 337 & 98 & 435 & 0 & 0 & 435 \\
        \hline
        MESSIDOR 2~\cite{decenciere2014feedback} & 1287 & 457 & 1744 & 0 & 0 & 1744 \\
        \hline
        DIARETDB1~\cite{kauppi2007diaretdb1} & 43 & 46 & 89 & 0 & 0 & 89 \\
        \hline
        \hline
        Martínez (private) & 454 & 30 & 484 & 0 & 0 & 484 \\
        \hline
        HEC (private) & 26 & 9 & 35 & 0 & 0 & 35 \\
        \thickhline
        \textbf{Total} & 89433 & 27735 & 117168 & 39592 & 9569 & 61526 \\
        \thickhline
    \end{tabular}}
    \label{tab:datasets}
\end{table}

Contrast enhancement, on the other hand, has been widely applied since Graham's contribution to Kaggle EyePACS competition~\cite{graham2015kaggle}. As the author originally mentioned, it facilitates to identify small red lesions associated with early DR grades, specially in areas of uneven illumination (Figures~\ref{fig:prepro-second} and \ref{fig:prepro-detail}). Formally, the contrast enhanced image $I_\text{ce}$ is obtained from the original image $I$ by doing:
\begin{equation}
I_\text{ce}(i,j;\sigma) = \alpha I(i,j) - \beta G(i,j;\sigma  \ast I(i,j) + \gamma,    
\label{eq:preprocessing}
\end{equation}
where $i,j$ are the horizontal and vertical coordinates of each pixel, $\alpha = \beta = 4$ are blending parameters, $\ast$ is the convolution operator, $G(\cdot)$ is a Gaussian kernel with $\sigma = \frac{w}{90}$ for $w$ the width of $I$, and $\gamma = 128$ is an anchor intensity. To avoid undesired artifacts when applying the kernel in the borders of the FOV, pixels outside it are replaced by the mean RGB intensities of the pixels within the convex hull of the originally predicted FOV (Figure~\ref{fig:prepro-second}, left). Notice from Eq.~\ref{eq:preprocessing} that the Gaussian filter approximates a background (Figure~\ref{fig:prepro-second}, center) that, once subtracted to the image, enhances in $I_\text{ce}$ (Figure~\ref{fig:prepro-second}, right) both tiny red lesions and thin vessels (Figure~\ref{fig:prepro-detail}).

\subsection{Deep learning model and training setup}

\begin{figure}
    \centering
    \includegraphics[width=0.9\textwidth]{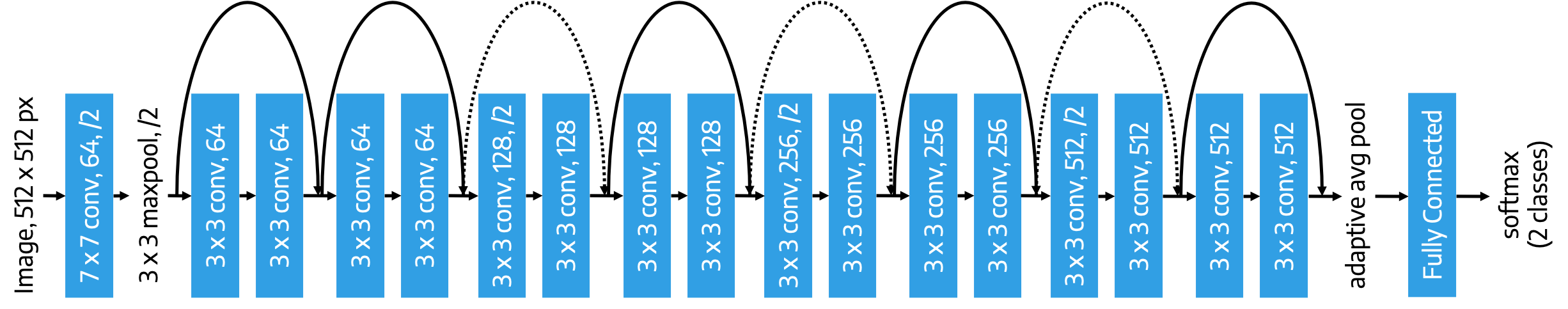}
    \vspace{0.3cm}
    \caption{ResNet-18 architecture of our proposed baseline. Curved arrows represents standard residual connections, while dotted ones are residual connections with zero-padding to match dimensions.}
    \label{fig:schematic-network}
\end{figure}

We chose a ResNet~\cite{he2016deep} as neural network backbone due to its popularity for referable DR classification~\cite{bellemo2019artificial,liu2019referable,hacisoftaoglu2020deep,yip2020technical,dai2021deep}. 
Among its many variants, we selected a ResNet-18 (Figure~\ref{fig:schematic-network}), the one with the smallest capacity. No adaptations were made on the architecture compared to its original definition.~\cite{he2016deep}
Inputs are RGB fundus images preprocessed as described in Section~\ref{subsec:methods-preprocessing}, resized to $512 \times 512$ pixels and with intensities normalized to the interval $[-1, +1]$. 
Outputs, on the other hand, are one-hot encoding vectors for non-referable (0) and referable (1) DR classes.
The network comprises 17 convolutional layers and a fully connected one.
Each convolutional layer is followed by batch normalization and ReLU activations, with a number of filters that is doubled every 4 layers, starting with 64. 
Adaptive average pooling is used to map the last tensor of convolutional activations into a vector of fixed length 512 features to feed the last classification layer.
A softmax activation is applied on top of the output logits to map them into probabilities.

Images from databases described in Section~\label{subsec:materials} were used to construct our training, validation and test partitions (Table~\ref{tab:datasets}). 
As EyePACS~\footnote[2]{\url{https://www.kaggle.com/c/diabetic-retinopathy-detection/data}} and IDRiD~\cite{porwal2018indian} only provide training and test splits, we integrated 10\% of their corresponding training sets in our validation set, preserving the remaining ones for training and their test sets for testing. 
Similarly, we followed the pre-defined partitions from DDR~\cite{li2019diagnostic}, and the original DeepDRID~\cite{liu2022deepdrid} training set was used for training. 
As DeepDRID does not provide labels for its test set, we used its validation set for testing instead. 
Likewise, we discarded the validation and test sets from APTOS2019~\footnote[1]{\url{https://www.kaggle.com/competitions/aptos2019-blindness-detection/data}} and integrated only its training set to ours. The remaining databases (FCM-UNA~\cite{benitez2021dataset}, 1000Fundus~\cite{cen2021automatic}, DR2~\cite{pires2015beyond}, MESSIDOR 2~\cite{decenciere2014feedback} and DIARETDB1~\cite{kauppi2007diaretdb1}) were all used as test sets.
Since 1000Fundus includes images with multiple conditions, we created a test set using only normal subjects and those with different DR grades, discarding other diseases and observations.
Finally, Mart\'inez and HEC sets were used as independent test sets, too.

We used the pretrained on ImageNet, PyTorch 1.10 
implementation of this network. 
A standard cross-entropy loss with no class weighting was used as training objective, minimized using Adam optimization with an initial learning rate of 0.0001.
This parameter was decreased to its half every time the area under the receiver-operating characteristic curve (AUC) in the validation set was not improved for a maximum of 20 epochs.
Batch size was fixed to 128 images per iteration, training for a maximum of 150 epochs and using early stopping with a patience of 40 epochs, based in the validation set AUC. 
Additional regularization was introduced using weight decay with a rate of 0.001 and online data augmentation techniques that were randomly and successively applied on the images before contrast enhancement.
Operations included color alterations, horizontal and vertical flippings, rotations and scalings, with parameters empirically fixed using the validation set.
The model was trained on a cloud-computing platform using an NVIDIA V100 GPU with 16 GBs of RAM, and evaluated in a standard i7 CPU with 16 GB of RAM and an NVIDIA 3060 GPU with 12 GB of memory.

\subsection{Evaluation metrics and data analysis}

We quantitatively evaluated the proposed baseline using standard binary classification metrics such as AUC, sensitivity (Se) and specificity (Sp). 
AUC was computed directly from referable DR probabilities predicted by the model. 
Se and Sp were calculated using the argmax of these outputs, hence without choosing an optimal operating point. 
Statistically testing the differences in AUC is not possible as models included in the comparison did not release their individual predictions.
Therefore, we used 95\% confidence intervals (95\% CI) for all metrics, obtained using bootstrap with replacement with $n = 1000$ samples for ROC computation and Wilson's method~\cite{wilson1927probable} for Se and Sp.
To qualitatively analyze results, class activation maps were automatically obtained using the XGrad-Cam method~\cite{fu2020axiom}, as implemented in Torchcam~\footnote[3]{\url{https://frgfm.github.io/torch-cam/}}. 
To account for multiscale features, the algorithm was applied to the second, third and fourth convolutional blocks of the network, and fused into a single one. 
In each case, the maps were computed for the predicted class: thus, if the model predicted a non-referable case, then attributions were obtained for that specific class, and viceversa. 
We also computed the uncertainty of the model as the entropy of the predicted probabilities to study how much it varies under different scenarios.

\section{RESULTS AND DISCUSSION}
\label{sec:results}

Table~\ref{tab:results-and-sota} presents the AUC, Se and Sp values and their 95\% CI obtained for classifying referable DR cases in each test set. Results reported by other deep learning models evaluated with the same data sets and metrics than ours are also included.
To the best of our knowledge, existing works only report comparable values in DDR, DR2, EyePACS, IDRiD and MESSIDOR 2, while ours is the first to be also evaluated in 1000Fundus, DeepDRID, DIARETDB1 and FCM-UMA.
When mixing all public datasets, our ResNet reported an AUC = 0.955 (0.953 - 0.956), with Se = 0.752 (0.744 - 0.759) and Sp = 0.977 (0.976 - 0.979).
In terms of computational efficiency, each prediction took 0.07 $\pm$ 0.004 seconds per image. 
In our private data sets, the model showed results comparable to those in the public ones, including in Martínez, which has images acquired with a device unseen during training.
Surprisingly, our results in public sets are in pair or even better than those reported in the literature, which lie within or below the 95\% CIs. 
In DDR, our ResNet obtained AUC values higher than those achieved with the lesion detection model by Zago \textit{et al.}~\cite{zago2020diabetic}. 
A similar behavior is observed in IDRiD, were our model slightly outperforms Hervella \textit{et al.}~\cite{hervella2022multimodal} method as well. 
Notice that this latter model uses a CNN pretrained in a paired data set of co-registered CFPs and fluorescein angiographies, which are costly to obtain. 
Alternatively, our simple model fine-tuned from ImageNet was able to obtain comparable results. 
When training their approach using the same strategy than ours, Hervella \textit{et al.} reported numbers notoriously lower than those achieved with our method~\cite{hervella2022multimodal} (AUC = 0.885 vs. 0.946, respectively). 
In MESSIDOR 2, the highest observed AUC and Se values were reported by Gulshan \textit{et al.}~\cite{gulshan2016development}. 
However, such an approach was trained using 128175 images, while ours uses one-third of that amount. 
The method by Gargeya \textit{et al.}~\cite{gargeya2017automated}, on the other hand, which used 75137 images for training, reported lower AUC and Sp values than our ResNet, which was trained with half this amount of scans. 
This might be attributed to our learning scheme based on images collected from multiple sources and a well-calibrated data augmentation strategy, which allows to diversify the appearance of CFPs in the training set and therefore strengthen the model generalization ability. 
Li \textit{et al.}~\cite{li2022deep} achieved slightly better results using an ensemble of five Inception-v4 networks trained with less than 9000 images of size $299 \times 299$ pixels each. 
Instead, we used a larger training set with higher resolution images but for learning a single model, which in practice corresponds to a less number of parameters. 
Furthermore, authors reported an AUC $=0.951$ (0.947-0.954) when training a single model with $1488 \times 1488$ pixel images, which is lower than what we obtained. 
In the EyePACS set, which features more than 50000 studies for evaluation, our algorithm slightly exceeds the AUCs reported by Pires \textit{et al.}~\cite{pires2019data} and Quellec \textit{et al.}~\cite{quellec2017deep}. 
Finally, our model also obtained a higher AUC than Pires \textit{et al.}~\cite{pires2019data} in DR2. Such an approach uses data augmentation and multiple trainings at different image resolutions to make the model robust to changes in scaling. Ours, instead, achieves this ability by simply learning from diverse data.

Any automated screening system should alert the patient if dangerous DR signs are detected in any of their eyes. 
Thus, we conducted a per-patient evaluation combining responses from multiple images of the same patient, using EyePACS and MESSIDOR 2, which provide one image per eye of the same individual, and DeepDRID, which includes two images per eye.
A subject was considered referable if at least one of their images was labeled as that in the ground truth. 
Similarly, we took the maximum predicted probabilities of all images of their left and right eyes to produce a single probabilistic response of the model. 
Results are summarized at the bottom of Table~\ref{tab:results-and-sota}. 
When contrasting them with the per-image evaluation, improvements in Se are observed at the cost of decreases in Sp. 
These changes are more evident in DeepDRID and EyePACS, were Se is increased from 88.3\% to 98\% and from 73.2\% to 77.3\%, respectively, while Sp is decreased from 86.8\% to 80\% and from 97.9\% to 96.9\%, respectively. 
A similar yet less compelling behavior is also observed in MESSIDOR 2. 
AUC, on the other hand, is increased in DeepDRID when using multiple images per eye, while in EyePACS and MESSIDOR 2 the AUC is slightly slower. 
When comparing these values with other state-of-the-art alternatives, we observe that Pires \textit{et al.}\cite{pires2019data} reports higher results. 
However, tehir approach integrates responses from both eyes by feeding features from the classification network to a dedicated classifier. 
Ours, on the contrary, reproduces the alternative used by Zago \textit{et al.}~\cite{zago2020diabetic} of taking the maximum predicted probability. 
Hence, it is not possible to determine if these differences are due to the neural network itself or the patient-level classifier. 

\begin{table}[t]
    \centering
    \caption{Per-image (top) and per-patient (bottom) evaluation for referable DR classification in all test sets in terms of AUC, Se, Sp and their 95\% CI.}
    \vspace{0.3cm}
    \resizebox{0.8\textwidth}{!}{
    \begin{tabular}{ C{2.5cm} | C{4cm} | L{3.3cm} | L{3.3cm} | L{3.3cm} }
        \thickhline
        \multicolumn{5}{c}{\textbf{Per-image referable DR classification}} \\
        \thickhline
        \textbf{Test dataset}                                  & \textbf{Method}  & \textbf{AUC (95\% CI)} & \textbf{Se (95\% CI)} & \textbf{Sp (95\% CI)}  \\
        \thickhline
        1000Fundus   &  \textbf{ResNet-18 (ours)}   & \textbf{1.000} (1.000 - 1.000) & \textbf{1.000} (0.958 - 1.000) & \textbf{0.982} (0.906 - 0.997) \\
        \hline
        \hline
        \multirow{2}{2.5cm}{\centering DDR}  
            & Zago \textit{et al.} 2020 \cite{zago2020diabetic}       & 0.833 (0.819 - 0.846) & -                   & - \\  
        \cline{2-5}
            & \textbf{ResNet-18 (ours)}       & \textbf{0.965} (0.960 - 0.970) & \textbf{0.749} (0.728 - 0.769) & \textbf{0.978} (0.971 - 0.984) \\
        \hline
        \hline
        DeepDRID   &  \textbf{ResNet-18 (ours)}   & \textbf{0.959} (0.944 - 0.972) & \textbf{0.883} (0.828 - 0.922) & \textbf{0.868} (0.817 - 0.907) \\
        \hline
        \hline
        DIARETDB1   &  \textbf{ResNet-18 (ours)}   & \textbf{0.981} (0.956 - 0.999) & \textbf{0.957} (0.855 - 0.988) & \textbf{0.930} (0.814 - 0.976) \\
        \hline
        \hline        
        \multirow{2}{2.5cm}{\centering DR2}  
            & Pires \textit{et al.} 2019 \cite{pires2019data}      & 0.963 (0.938 - 0.981)    & -                   & - \\  
        \cline{2-5}
            & \textbf{ResNet-18 (ours)}       & \textbf{0.974} (0.962 - 0.985) & \textbf{0.847} (0.763 - 0.905) & \textbf{0.961} (0.935 - 0.977) \\
        \hline
        \hline     
        \multirow{3}{2.5cm}{\centering EyePACS}  
            & Quellec \textit{et al.} 2017 \cite{quellec2017deep}    & 0.944                    & -                   & - \\  
        \cline{2-5}
            & Pires \textit{et al.} 2019 \cite{pires2019data}     & 0.946                    & -                   & - \\  
        \cline{2-5}
            & \textbf{ResNet-18 (ours)}       & \textbf{0.951} (0.949 - 0.954) & \textbf{0.732} (0.723 - 0.740) & \textbf{0.979} (0.978 - 0.980) \\
        \hline
        \hline     
        FCM-UNA   &  \textbf{ResNet-18 (ours)}   & \textbf{0.986} (0.980 - 0.992) & \textbf{0.882} (0.852 - 0.906) & \textbf{0.990} (0.963 - 0.997) \\
        \hline
        \hline        
        \multirow{3}{2.5cm}{\centering IDRiD}  
            & Zago \textit{et al.} 2020 \cite{zago2020diabetic}      & 0.796 (0.715 - 0.892)    & -                & - \\
        \cline{2-5}
            & Hervella \textit{et al.} 2022 \cite{hervella2022multimodal}   & 0.944                    & -                & - \\
        \cline{2-5}
            & \textbf{ResNet-18 (ours)}       & \textbf{0.949} (0.914 - 0.980)    & \textbf{0.828} (0.718 - 0.901)   & \textbf{0.897} (0.764 - 0.959) \\
        \hline
        \hline
        \multirow{5}{2.5cm}{\centering MESSIDOR 2}  
            & Gulshan \textit{et al.} 2016 \cite{gulshan2016development}    & \textbf{0.990} (0.986 - 0.995)    & \textbf{0.961} (0.924 - 0.983)   & 0.939 (0.924 - 0.953) \\  
        \cline{2-5}
            & Gargeya \textit{et al.} 2017 \cite{gargeya2017automated}   & 0.940                    & 0.930                   & 0.870 \\ 
        \cline{2-5}
            & Voets \textit{et al.} 2019 \cite{voets2019reproduction}      & 0.853 (0.835 - 0.871)    & 0.818                   & 0.687 \\ 
        \cline{2-5}
        & Zago \textit{et al.} 2020 \cite{zago2020diabetic}      & 0.944 (0.925 - 0.966)    & 0.900 (0.860 - 0.961)   & 0.870 (0.863 - 0.871) \\  
        \cline{2-5}
        & Li \textit{et al.} 2022 \cite{li2022deep}     & 0.977 (0.974 - 0.981)     & 0.923 (0.917 - 0.925)     & 0.947 (0.937 - 0.954) \\
        \cline{2-5}
            & \textbf{ResNet-18 (ours)}       & 0.973 (0.967 - 0.979)    & 0.895 (0.863 - 0.920)   & \textbf{0.941} (0.927 - 0.953) \\
        \hline
        \hline
        HEC   &  \textbf{ResNet-18 (ours)}   & \textbf{0.961} (0.900 - 1.000) & \textbf{1.000} (0.610 - 1.000) & \textbf{0.862} (0.694 - 0.945) \\
        \hline
        \hline
        Martínez   &  \textbf{ResNet-18 (ours)}   & \textbf{0.955} (0.927 - 0.980) & \textbf{0.800} (0.627 - 0.905) & \textbf{0.934} (0.907 - 0.953) \\
        \thickhline 
        \multicolumn{5}{c}{} \\
        \thickhline 
        \multicolumn{5}{c}{\textbf{Per-patient referable DR classification}} \\
        \thickhline
        \multirow{2}{2.5cm}{\centering DeepDRID \\ (100 patients)}   & \multirow{2}{4cm}{\centering \textbf{ResNet-18 (ours)}}  
          & \multirow{2}{3.3cm}{\textbf{0.980} (0.957 - 0.999)} & \multirow{2}{3.3cm}{\textbf{0.980} (0.895 - 0.996)} & \multirow{2}{3.3cm}{\textbf{0.800} (0.670 - 0.888)} \\
          & & & & \\
        \hline
        \hline
        \multirow{3}{2.5cm}{\centering EyePACS \\ (26788 patients)}  
            & Pires \textit{et al.} 2019 \cite{pires2019data}     & \textbf{0.955} (0.951 - 0.958)                   & -                   & - \\  
        \cline{2-5}
            & Zago \textit{et al.} 2020 \cite{zago2020diabetic}      & 0.821 (0.812 - 0.829)    & -   & - \\  
        \cline{2-5}
            & \textbf{ResNet-18 (ours)}       & 0.948 (0.945 - 0.951) & \textbf{0.773} (0.762 - 0.783) & \textbf{0.969} (0.966 - 0.971) \\
        \hline
        \hline
        \multirow{3}{2.5cm}{\centering MESSIDOR 2 \\ (870 patients)}  
            & Pires \textit{et al.} 2019 \cite{pires2019data}    & \textbf{0.982} (0.974 - 0.989)    & -   & - \\  
        \cline{2-5}
            & Zago \textit{et al.} 2020 \cite{zago2020diabetic}      & 0.944 (0.927 - 0.965)    & -   & - \\  
        \cline{2-5}    
            & \textbf{ResNet-18 (ours)}       & 0.970 (0.960 - 0.979)    & 0.907 (0.866 - 0.936)   & \textbf{0.920} (0.896 - 0.939) \\
        \thickhline
    \end{tabular}}
    \label{tab:results-and-sota}
\end{table}

\begin{figure}[t]
  \centering
  \begin{subfigure}[t!]{0.32\textwidth}
  \includegraphics[width=\textwidth]{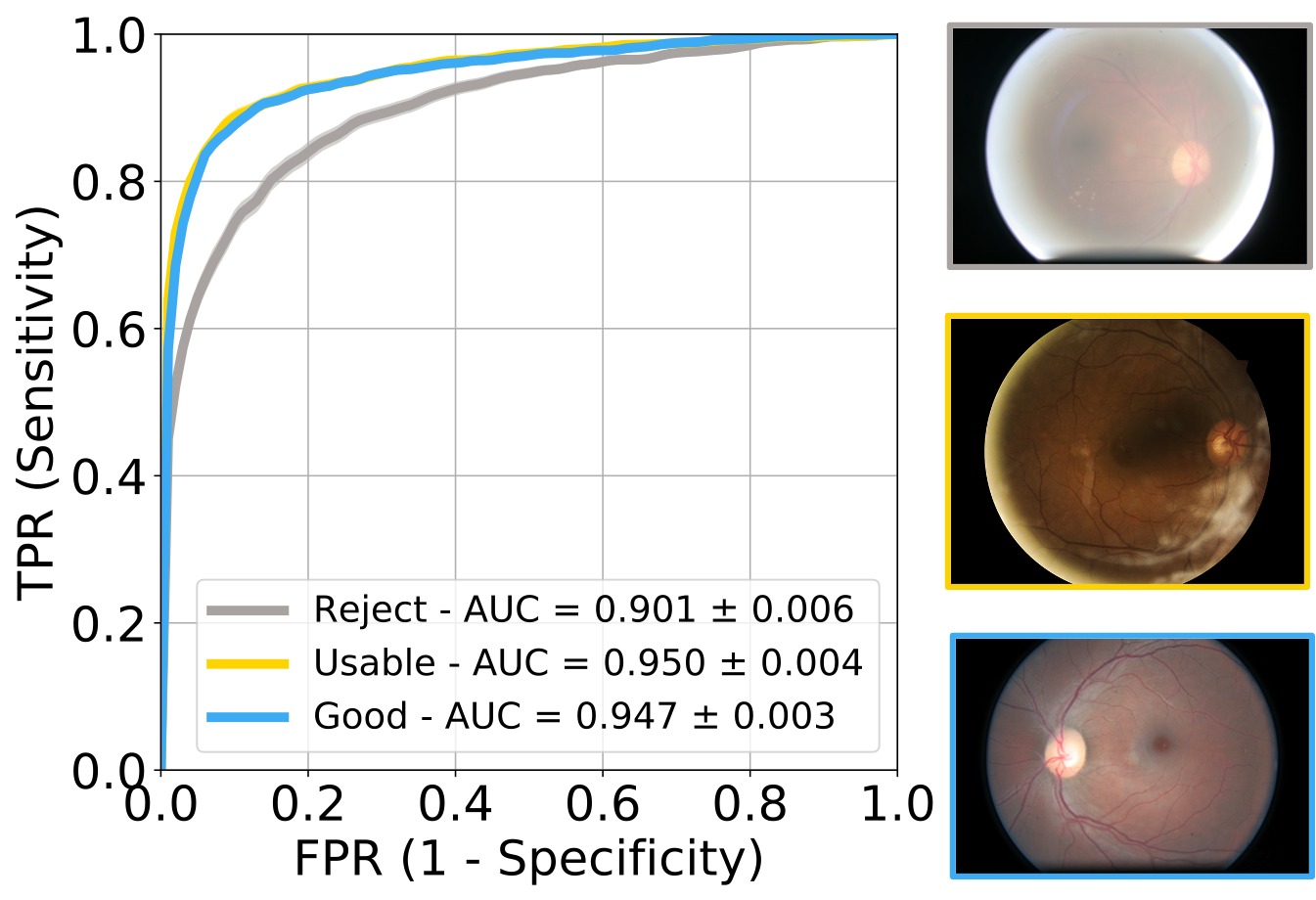}\label{fig:quality-examples-eyepacs-fu}
  \caption{EyePACS (Fu \textit{et al.}~\cite{fu2019evaluation} labels)}
  \end{subfigure}
  \begin{subfigure}[t!]{0.32\textwidth}
  \includegraphics[width=\textwidth]{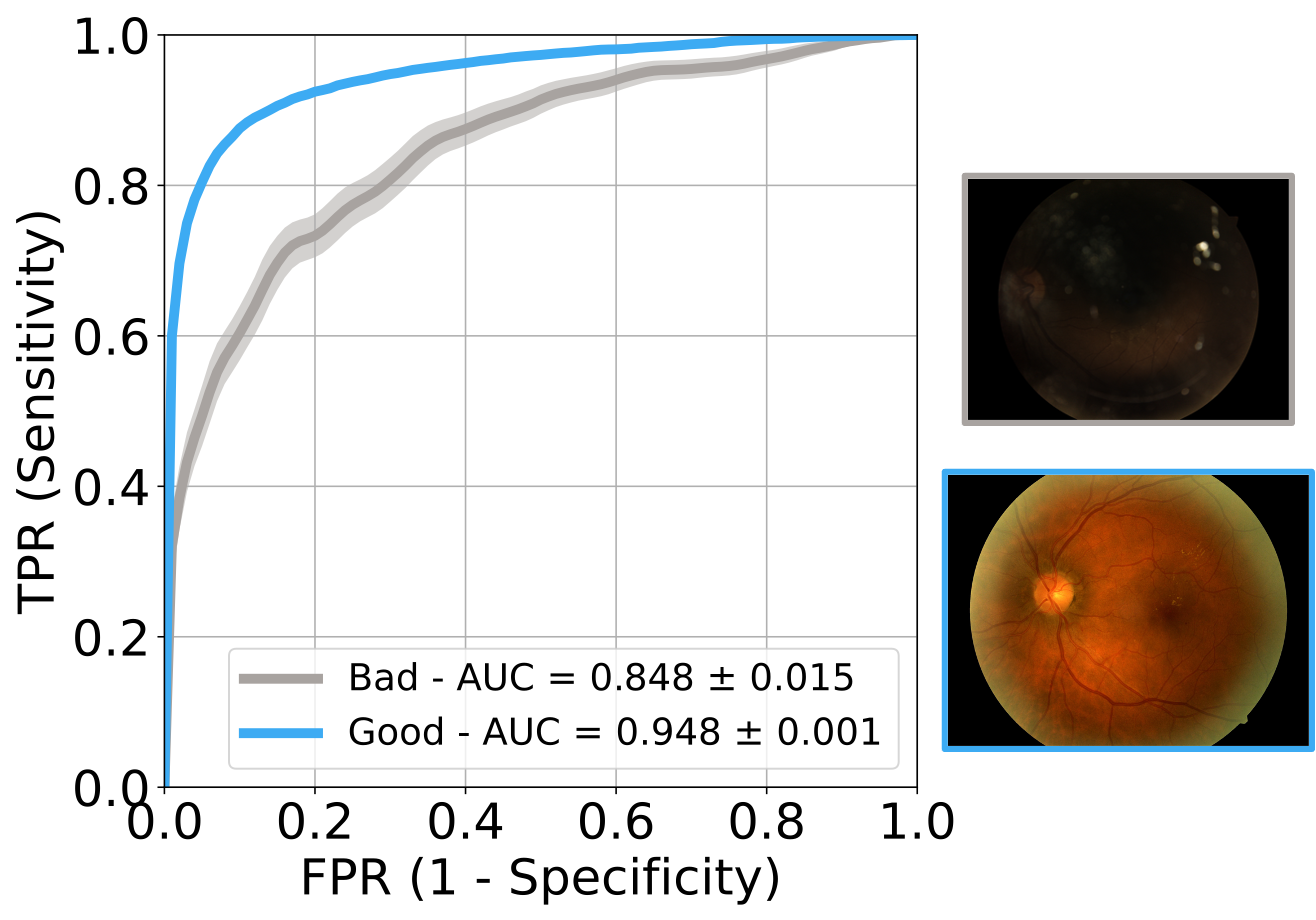}\label{fig:quality-examples-eyepacs-zhou}
  \caption{EyePACS (Zhou \textit{et al.}~\cite{zhou2018fundus} labels)}
  \end{subfigure}
  \begin{subfigure}[t!]{0.33\textwidth}
  \includegraphics[width=\textwidth]{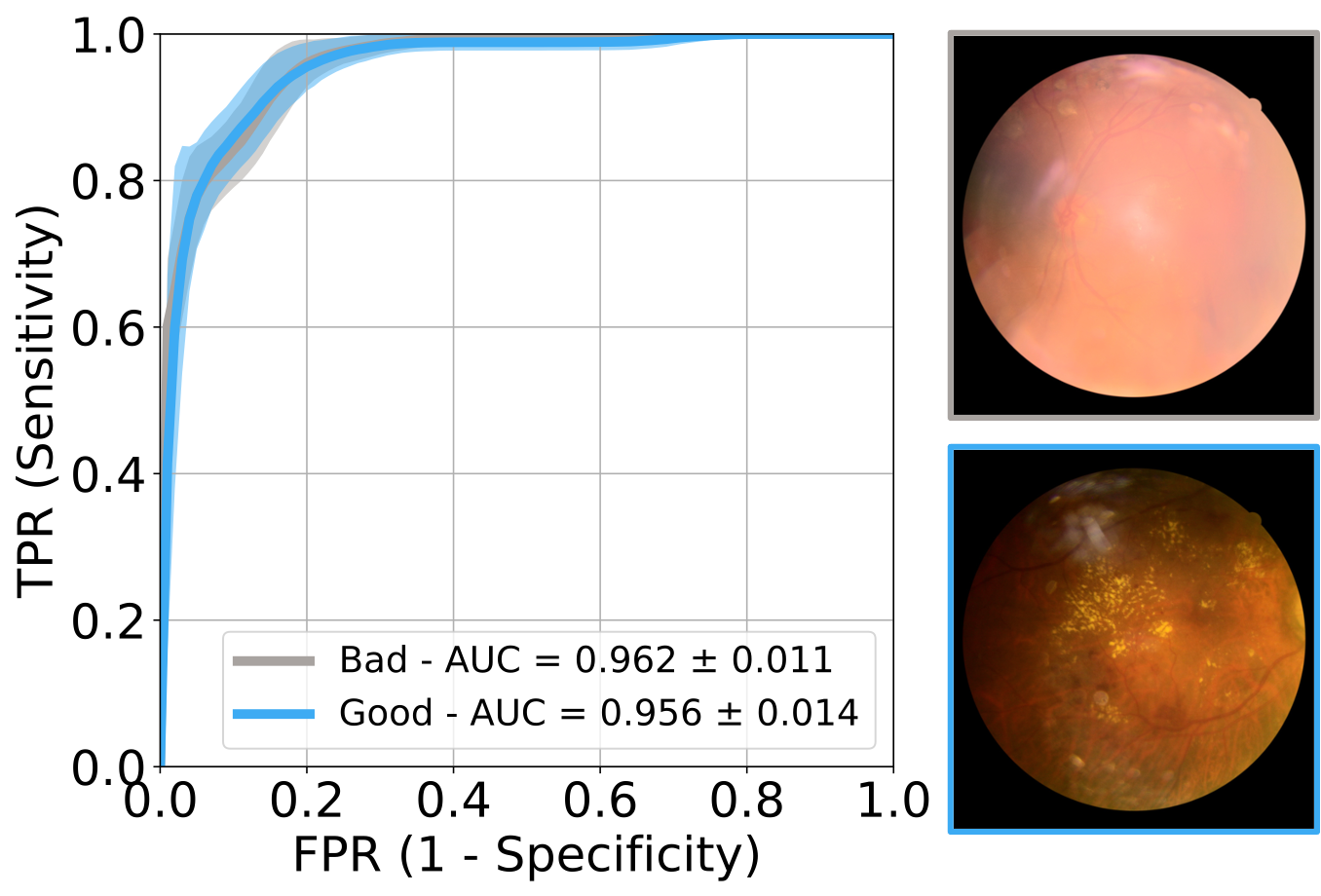}\label{fig:quality-examples-deepdrid}
  \caption{DeepDRID}
  \end{subfigure}
  \vspace{0.3cm}
  \caption{ROC curves and AUC values obtained for referable DR detection in images separated by their quality. Shadings correspond to 95\% CI. Borders of each image are color coded according to their quality label.}
\label{fig:image-quality}
\end{figure}

We performed an additional experiment discriminating classification outputs according to quality labels provided for EyePACS by Fu \textit{et al.}~\cite{fu2019evaluation} ("reject", "usable" and "good") and Zhou \textit{et al.}~\cite{zhou2018fundus} ("bad" and "good"), and those available for DeepDRID ("bad" and "good"). 
Figure~\ref{fig:image-quality} presents the ROC curves obtained for each type of image, jointly with image samples for each category. 
In EyePACS, the model obtained lower AUC values for bad quality images, regardless of the source of the quality labels. 
On the other hand, results for usable and good images are almost equivalent when separated using Fu \textit{et al.}~\cite{fu2019evaluation} labels. 
Finally, a similar behavior is observed in DeepDRID, with AUC values that are comparable between good and bad quality scans.

\begin{figure}[t]
  \centering

  \begin{subfigure}[t!]{0.3\textwidth}
  \includegraphics[width=\textwidth]{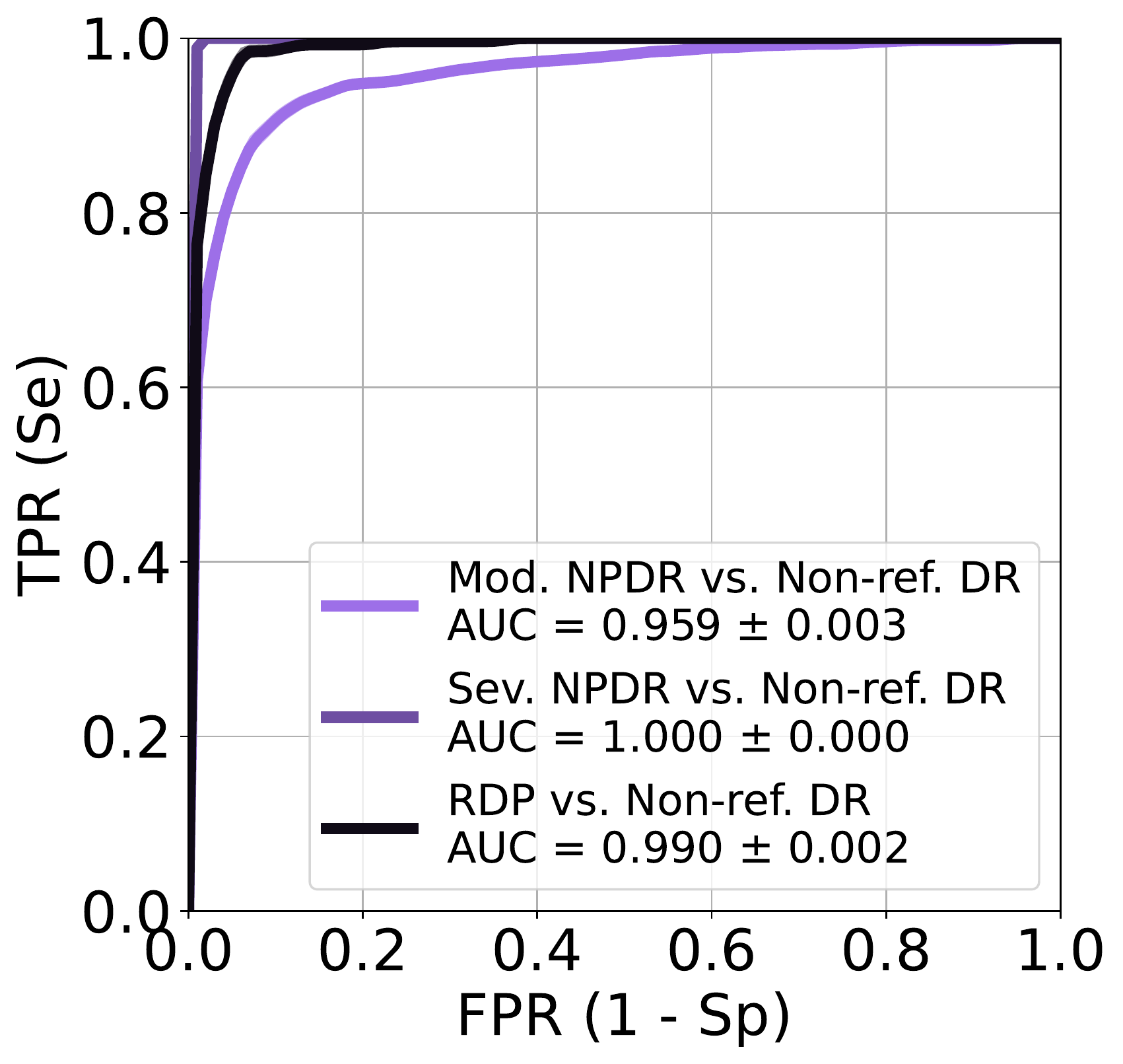}\label{fig:roc-ddr-grade}
  \caption{DDR}
  \end{subfigure}
  \begin{subfigure}[t!]{0.3\textwidth}
  \includegraphics[width=\textwidth]{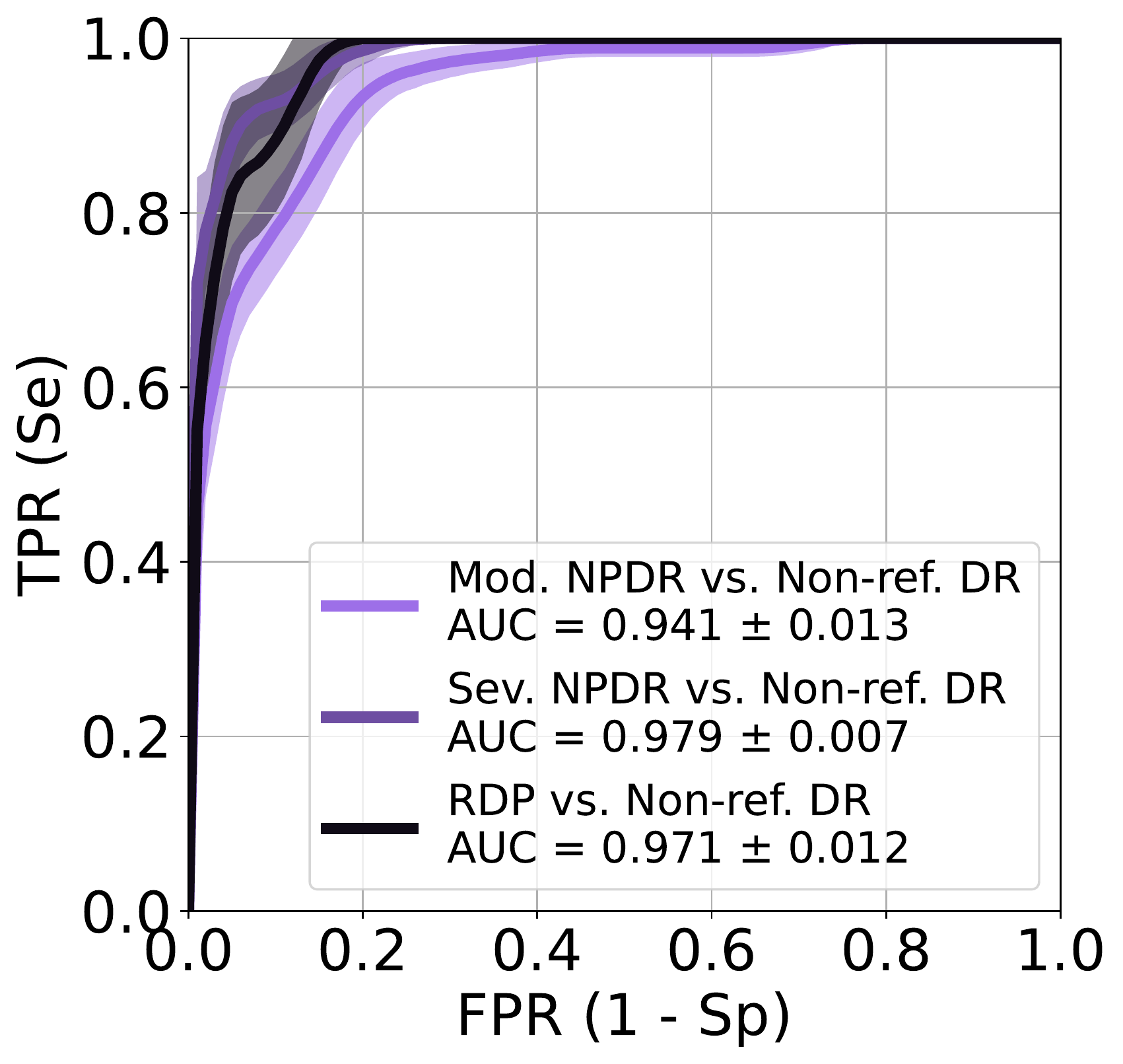}\label{fig:roc-deepdrid-grade}
  \caption{DeepDRID}
  \end{subfigure}
  \begin{subfigure}[t!]{0.3\textwidth}
  \includegraphics[width=\textwidth]{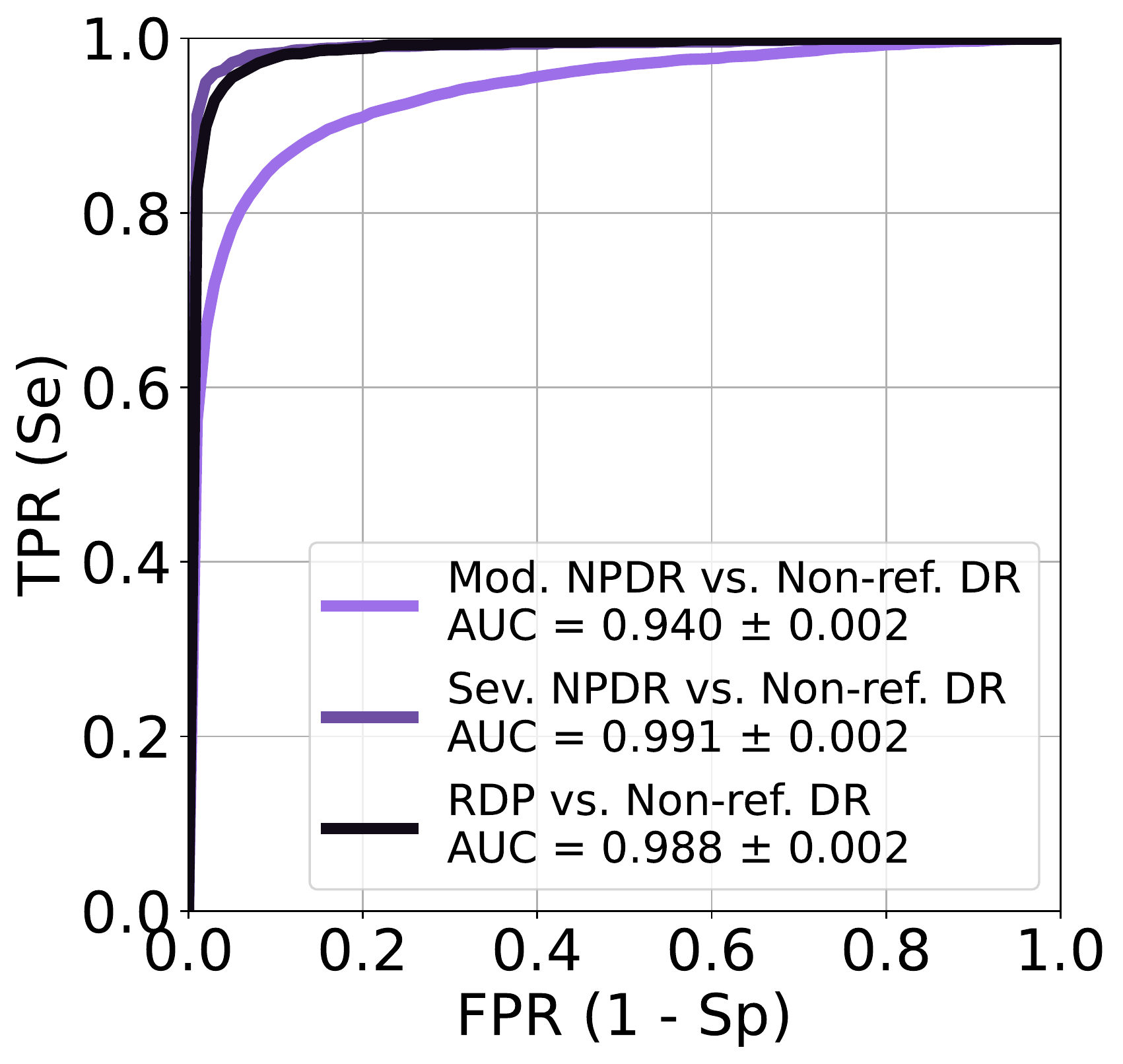}\label{fig:roc-eyepacs-grade}
  \caption{EyePACS}
  \end{subfigure}
  
  \vspace{0.2cm}
  \begin{subfigure}[t!]{0.3\textwidth}
  \includegraphics[width=\textwidth]{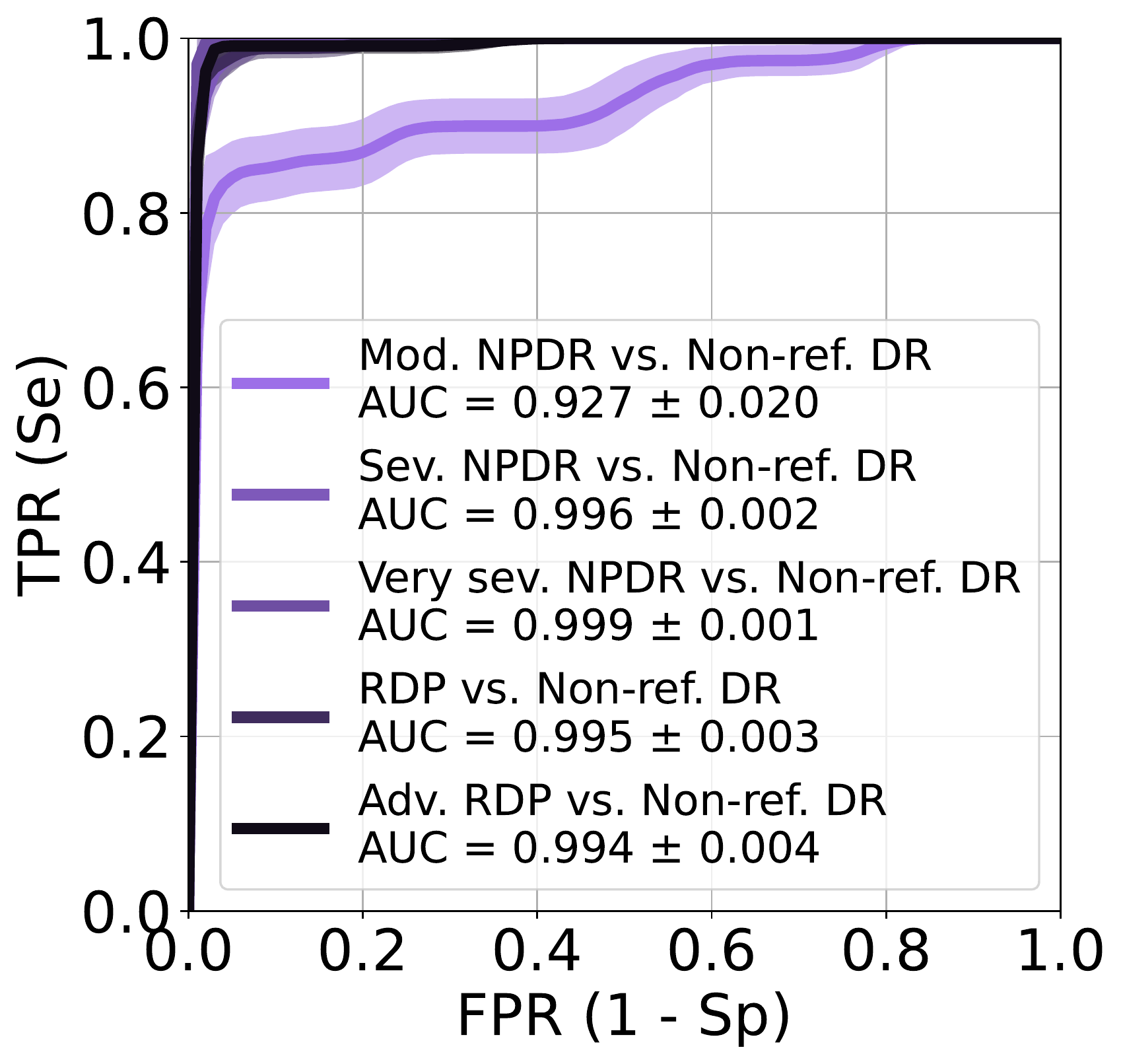}\label{fig:roc-fcm-grade}
  \caption{FCM-UNA}
  \end{subfigure}
  \begin{subfigure}[t!]{0.3\textwidth}
  \includegraphics[width=\textwidth]{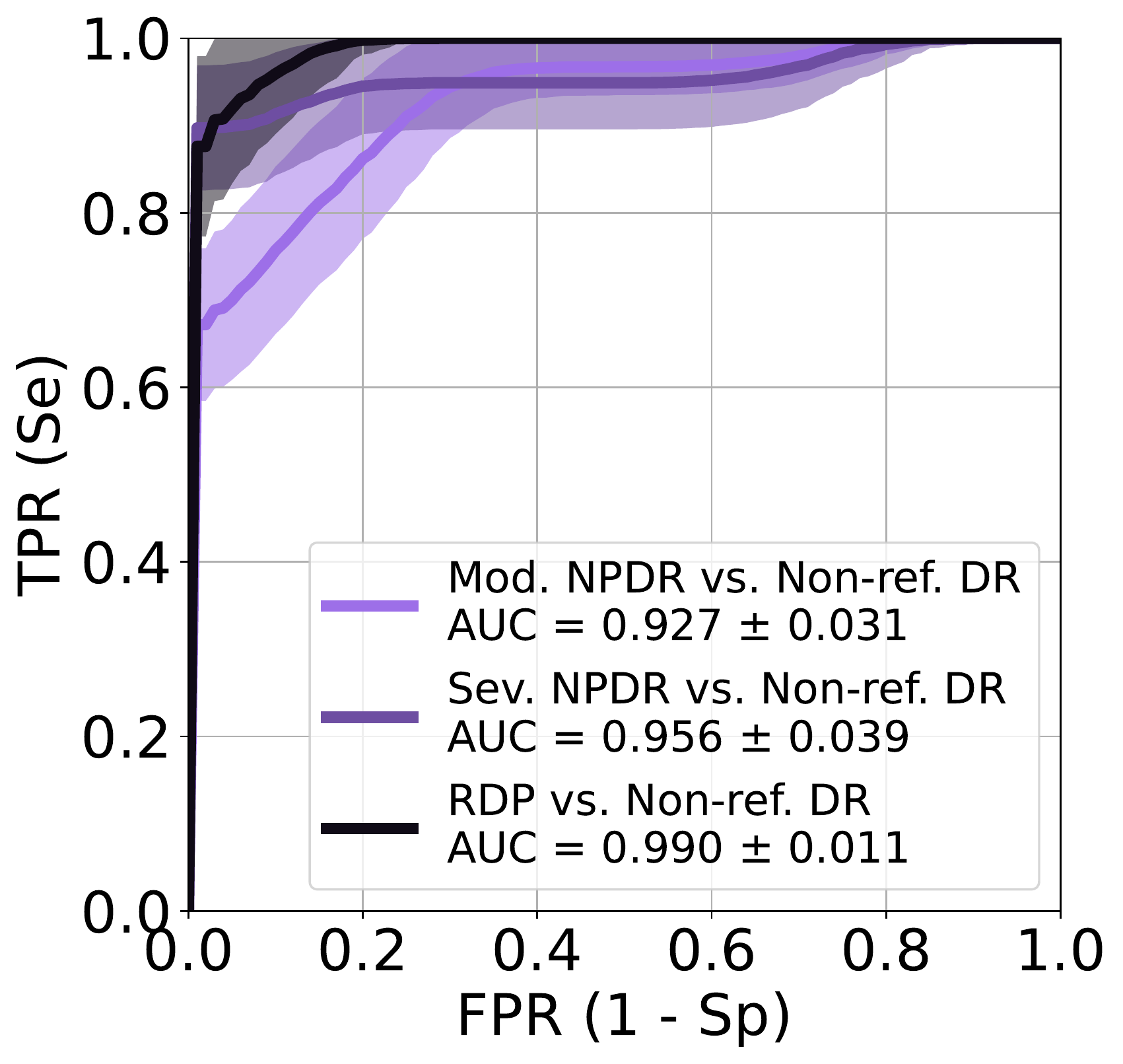}\label{fig:roc-idrid-grade}
  \caption{IDRiD}
  \end{subfigure}
  \begin{subfigure}[t!]{0.3\textwidth}
  \includegraphics[width=\textwidth]{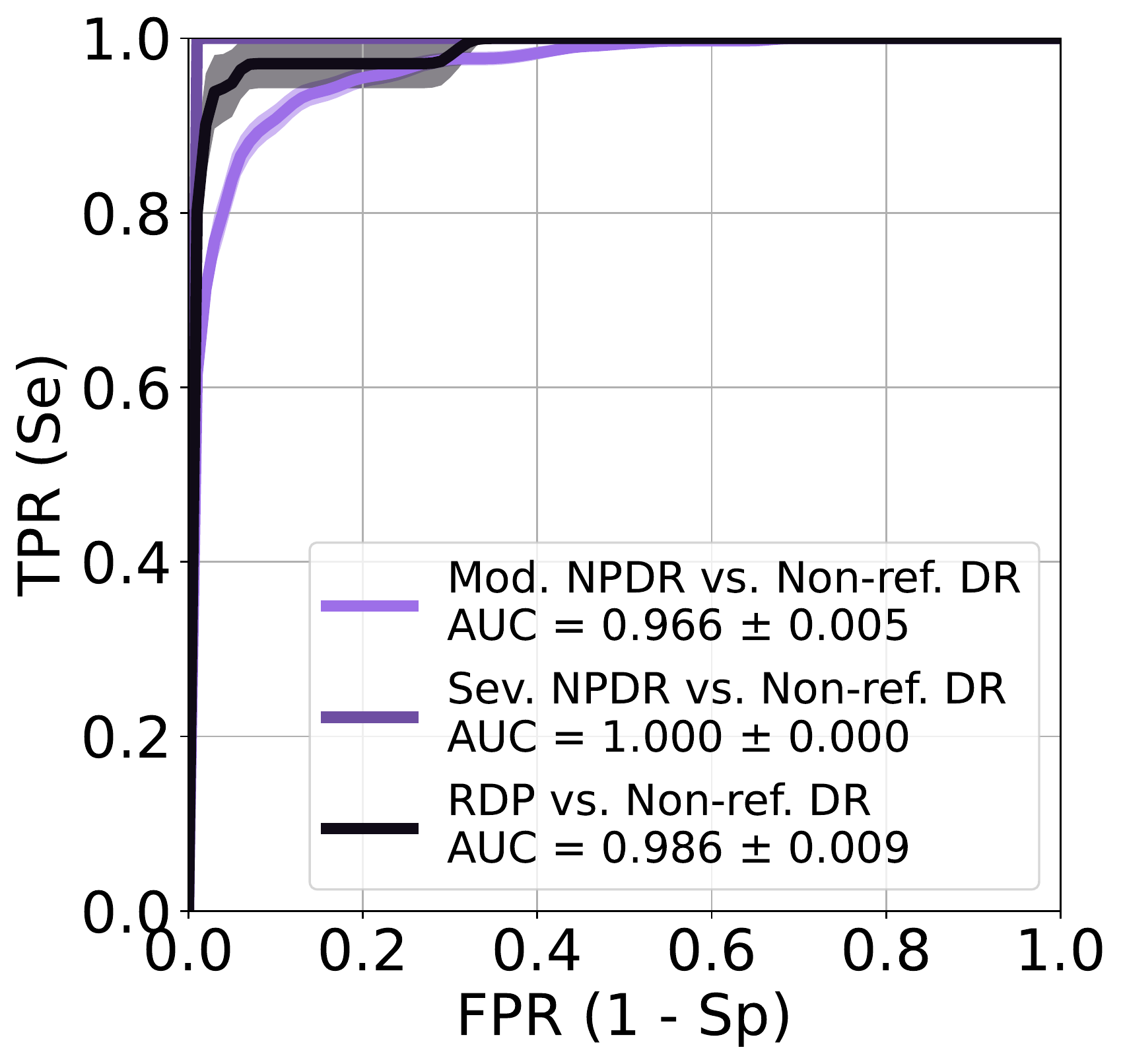}\label{fig:roc-messidor-grade}
  \caption{MESSIDOR 2}
  \end{subfigure}
  
  \vspace{0.3cm}
  \caption{ROC curves and AUC values obtained using referable DR probabilities to classify non-referable DR from moderate, severe and very severe NPDR, and RDP and advanced RDP. Shadings correspond to 95\% CI.}
\label{fig:roc-curves-per-grade}
\end{figure}

Figure~\ref{fig:roc-curves-per-grade} depicts ROC curves presenting the differences in performance to recognize cases of non-referable DR from moderate NPDR, first, then non-referable RD vs. severe NPDR, and finally non-referable RD from proliferative RD (PDR). This was achieved by separating subsets from DDR, IDRiD, MESSIDOR 2, EyePACS, DeepDRID and FCM-UMA using their disease grade, and creating different mixed versions in which the non referable set was always fix and the referable one changed. The highest AUC values were obtained for the more advanced grades, while results for moderate NPDR were also high yet smaller. Noticeable, in some cases RDP detection was less accurate than severe NPDR detection, although differences are not compelling.

\begin{figure}[t]
  \centering

  
  
  
  \includegraphics[width=0.95\textwidth]{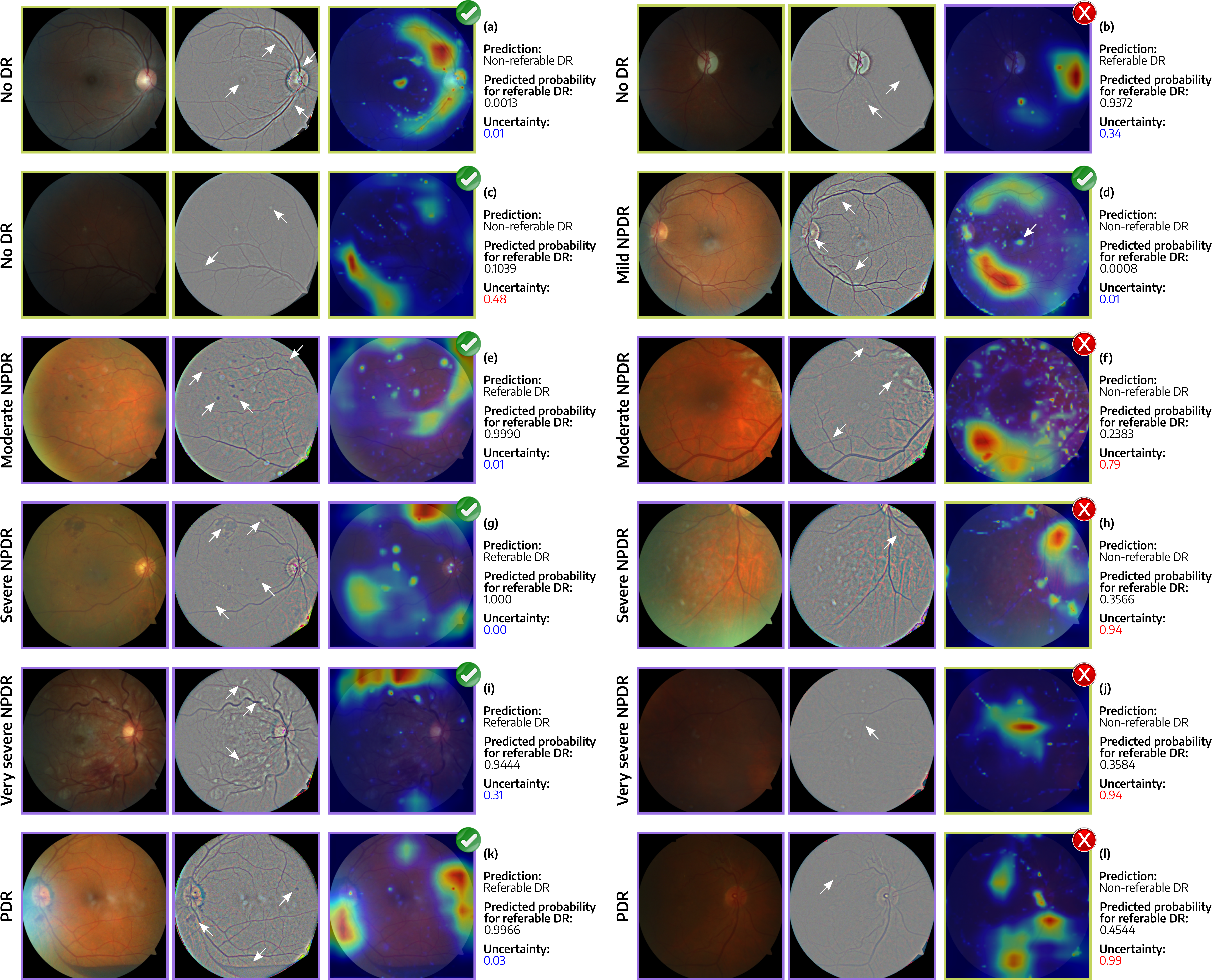}\label{fig:cam}
  \vspace{0.3cm}
  \caption{Qualitative results obtained in FCM-UNA dataset. Each example includes DR grade, the image cropped around the FOV area (left), its preprocessed version (center) and the associated XGrad-Cam (right), jointly with the predicted class, the predicted probability for referable DR and the uncertainty of the prediction.}
\label{fig:qualitative-cams}
\end{figure}

Finally, Figure~\ref{fig:qualitative-cams} illustrates some representative qualitative results obtained in images from FCM-UNA. Each example includes the image, its preprocessed version and its XGrad-Cam on top of the original scan. 
We also included the ground truth DR grade, a mark indicating if the image was correctly classified (\cmark) or not (\xmark), the predicted class and referable DR probability and the uncertainty estimate. 
According to the ground truth labels, the ResNet-18 model correctly classified almost all non referable DR cases in FCM-UNA, even in dark, low quality images as those shown in (a) and (c). 
XGrad-Cams in these cases highlight either the macula, the optic disc, the normal nerve fiber layers in the vascular arcades and some arteriovenous crossings (see arrows in (a) and (d)). Regions with cotton wool spots or even with no apparent clinical meaning are also activated in (c) (top and bottom arrows, respectively).
The case shown in (b) corresponds to one of the two false positives in FCM-UNA, which was classified as referable with a high probability. The image shows signs of hypertensive retinopathy, with an inferior ghost vessel and pigmented lesions at a macular level. The XGrad-Cam shows activation on these lesions, indicating that they were took into account for predicting the case as referable.
Alternatively, (d) depicts an image indicated as non-referable DR by the ground truth labeling, that the model correctly classified as such. Lesions associated with a maculopathy are observed at the vecinity of the fovea, including a potential exudate, that the activation map seems to ignore. 
When analyzing the referable DR examples, we can observe that the model has usually high confidence when classifies them correctly, which is reflected by the low uncertainty values. 
Although activation maps localize certain microhemorrhages (see arrows in (e), (g) and (k)) and exudates (top right arrow in (g) and (i)), they usually ignore areas with large hemorrhages (top left arrow in (g), bottom arrows in (i) and (k)).
Furthermore, big activation blobs are seen in (e), (g), (j) and (i) over areas that do not correspond to relevant signs but e.g. artifacts (j). 
The errors on the right hand side of Figure~\ref{fig:qualitative-cams}, on the other hand, are associated to low quality inputs in which the model made predictions with high uncertainty. 
The case in (f), for instance, is a false negative of the model, even though the image shows pathological signs such as microhemorrhages, soft exudates, venous dilations and pathological arteriovenous crossings. The XGrad-Cam only highlights the latter, which might indicate that the model struggled to identify the other signs and therefore made a negative prediction.
A similar behavior is seen in (l), where only a small bright lesion is captured as relevant. 
The model detected cases (h) and (j) as non-referable DR, while they are labeled as severe NPDR cases in FCM-UNA. However, no abnormalities are clearly observable, which might indicate that they are wrongly tagged in the database.
Finally, we asked two experienced ophthalmologists (ML and MM) to evaluate the utility of the contrast enhancement operation. According to them, it allows to better visualize vascular structures (b, j), white lesions (c, l) and small hemorrhages (e, f, g, k), although in other cases also enhances artifacts (i), does not increase the overall quality of the image (a, d) or even hide large hemorrhages (bottom arrow in (i)).

\section{CONCLUSIONS}
\label{sec:conclusions}

We presented a strong baseline for referable DR detection from CFPs based on a standard ResNet-18 classification network. 
By combining data from multiple sources and a calibrated data augmentation strategy, we showed that it is possible to obtain an accurate model that performs in line or even better than other more complex state-of-the-art approaches. 
This is in line with other studies showing that optimally trained baselines can be better than most of the recently published alternatives.~\cite{goyal2021revisiting} 
Furthermore, it highlights the importance of collecting more diverse data before extensively modifying the methodology.
In particular, our model showed high AUC values even when applied on our in-house databases, highlighting its generalization ability to inputs acquired with previously unseen devices, most likely due to uniforming images using a common preprocessing approach.
By reproducing multiple clinically relevant scenarios, our evaluation simultaneously showed that the model can be used as a strong baseline but some weak points should be improved, aiding us to envision other promising lines of research. 
In particular, we observed that our ResNet struggles more to recognize moderate NPDR cases than those suffering from more advanced disease grades. 
While this might be a consequence of potential ambiguities in the ground truth labels, future efforts should be made in improving the discrimination ability in these cases.
One potential way to do so is by training a multitask version that simultaneously predict the referable DR probability and the presence/absence of DR related lesions such as microaneurysms, hemorrhages, exudates or neovascularizations. 
The complementary nature of these tasks might aid to further improve in previously ambiguous classification scenarios.
Moreover, we qualitatively observe that the contrast equalization technique, despite useful to highlight small lesions, might enhance acquisition artifacts and hide large hemorrhages. Learning preprocessing parameters or simply creating a multichannel input by combining this image with the original one might eventually solve this limitation and let the network chose relevant features by its own. 
Furthermore, a certain relationship seems to arise between the image quality or the correctness of the response with prediction uncertainties. This is aligned with other previous observations in DR grading~\cite{araujo2020dr}, suggesting that this link should be further exploited to improve the accuracy of the models also in this binary task.
XGrad-Cams, on the other hand, showed a limited ability to detect some clinically relevant features such as exudates in the macular area. This can be associated with how the referable DR class was defined. Recent studies have suggested to integrate the risk of macular edema in this binary target~\cite{bellemo2019artificial}, as it also corresponds to a referable condition. This might certainly aid and enrich the outputs of the model. Finally, predicting the presence of lesions as mentioned above might help to improve these maps, further strengthening their applicability for lesion detection. 
To account for future comparisons, our results are publicly released at \url{https://github.com/TomasCast/sipaim-2022-resnet}.

\acknowledgments 
 
This study was partially funded by PICTs 2019-00070 and startup 2021-00023 granted by Agencia I+D+i (Argentina) and by PIP 2021-2023 11220200102472CO from CONICET (Argentina). We also thank NVIDIA for granting cloud-based GPU computing hours as part of their Applied Research Accelerator Program.

\bibliography{report} 
\bibliographystyle{spiebib} 

\end{document}